\documentclass{article}
\usepackage{cite}
\usepackage{graphicx}
\usepackage{dcolumn}

\begin{document}

\date{}
\title{Alternative solution to a quantum-mechanical four-particle system in one
dimension}
\author{Francisco M. Fern\'{a}ndez\thanks{%
fernande@quimica.unlp.edu.ar} \\
%EndAName
INIFTA, DQT, Sucursal 4, C.C 16, \\
1900 La Plata, Argentina}
\maketitle

\begin{abstract}
We propose an alternative solution to a quantum-mechanical four-particle
system in one dimension with two- and three-particle interactions. The
solution of the eigenvalue equation in center-of-mass and Jacobi coordinates
is considerably simpler than a procedure proposed recently based on
spherical coordinates.
\end{abstract}

\section{Introduction}

\label{sec:intro}

In this paper we propose an alternative solution to an exactly-solvable
quantum-mechanical four-particle model in one dimension with two- and
three-particle interactions. This model was recently proposed by Bakhshi and
Khoshdooni (BK)\cite{BK21}. In order to solve the eigenvalue equation they
resorted to center-of-mass and ``Jacobi'' coordinates\cite{C71} in order to
remove the free motion of the center of mass and obtain a new operator with
discrete spectrum. BK showed that the resulting Hamiltonian is separable in
spherical coordinates and provided solutions for the radial and angular
parts of the eigenvalue equation.

In addition to providing the alternative solution just mentioned we carry
out an analysis of BK's results. In section~\ref{sec:model} we introduce the
model, in section~\ref{sec:BK_solution} we outline and analyze BK's results,
in section~\ref{sec:correct_sol} we rederive those results in a somewhat
clearer way, in section~\ref{sec:exact_sol} we solve the problem directly in
Jacobi coordinates and, finally, in section~\ref{sec:conclusions} we outline
the main results and draw conclusions.

\section{The model}

\label{sec:model}

BK proposed the following four-particle Hamiltonian in one dimension\cite
{BK21}

\begin{equation}
H=-\frac{1}{2}\sum_{i=1}^{4}\frac{\partial ^{2}}{\partial x_{i}^{2}}+\frac{%
\omega ^{2}}{8}\sum_{i<j}^{4}\left( x_{i}-x_{j}\right) ^{2}+\frac{g_{1}^{2}}{%
\left( x_{1}+x_{2}-2x_{3}\right) ^{2}},  \label{eq:H_4p}
\end{equation}
and argued that they chose $\hbar =2m=1$ which is obviously inconsistent
with the factor $1/2$ in the kinetic-energy term (see reference~\cite{F20}
for a pedagogical introduction to dimensionless quantum-mechanical
equations). The strength $g_{1}^{2}$ of the Wolfes' interaction\cite{W74} is
expected to be a real and positive model parameter. BK stated that the
spectrum of this operator is real which is not the case unless one removes
the motion of the center of mass.

By means of the change of variables $\tilde{x}_{i}=\sqrt{\omega }x_{i}$ one
can easily prove that $H\left( \omega ,g_{1}\right) =\omega H\left(
1,g_{1}\right) $ so that one can choose $\omega =1$ without loss of
generality (see reference~\cite{F20} for other cases in which the number of
model parameters can be reduced).

\section{BK's solution in spherical coordinates}

\label{sec:BK_solution}

In order to solve the Schr\"{o}dinger equation BK resorted to the
center-of-mass and translation-invariant Jacobi coordinates\cite{C71}
\begin{eqnarray}
X_{1} &=&\frac{1}{\sqrt{2}}\left( x_{1}-x_{2}\right) ,  \nonumber \\
X_{2} &=&\frac{1}{\sqrt{6}}\left( x_{1}+x_{2}-2x_{3}\right) ,  \nonumber \\
X_{3} &=&\frac{1}{\sqrt{12}}\left( x_{1}+x_{2}+x_{3}-3x_{4}\right) ,
\nonumber \\
X &=&\frac{1}{2}\left( x_{1}+x_{2}+x_{3}+x_{4}\right) ,
\label{eq:Jacobi_coord}
\end{eqnarray}
and the spherical coordinates
\begin{eqnarray}
X_{1} &=&r\sin \theta \cos \phi ,  \nonumber \\
X_{2} &=&r\sin \theta \sin \phi ,  \nonumber \\
X_{3} &=&r\cos \theta ,  \nonumber \\
0 &\leq &r<\infty ,\;0\leq \theta \leq \pi ,\;0\leq \phi \leq 2\pi .
\label{eq:spherical_coord}
\end{eqnarray}
Curiously, BK chose the obviously incorrect domains $0\leq \theta \leq 2\pi
,\;0\leq \phi \leq \pi $. After carrying out such transformations and
removing the free motion of the center of mass the Hamiltonian operator
becomes
\begin{eqnarray}
H &=&-\frac{1}{2r^{2}}\frac{\partial }{\partial r}r^{2}\frac{\partial }{%
\partial r}+\frac{\omega ^{2}}{2}r^{2}+\frac{K}{2r^{2}},  \nonumber \\
K &=&-\frac{1}{\sin \theta }\frac{\partial }{\partial \theta }\sin \theta
\frac{\partial }{\partial \theta }+\frac{F}{\sin ^{2}\theta },  \nonumber \\
F &=&-\frac{\partial ^{2}}{\partial \phi ^{2}}+\frac{g_{1}^{2}}{3\sin
^{2}\phi }.  \label{eq:H_spherical}
\end{eqnarray}

In order to solve the eigenvalue equation for $H$ BK resorted to the method
of separation of variables and, consequently, chose a solution of the form
\begin{equation}
\psi (r,\theta ,\phi )=R(r)\Theta (\theta )\Phi (\phi ).
\label{eq:psi_product}
\end{equation}
However, they assumed that $\Theta (\theta )$ is a solution to $K\Theta
_{l}(\theta )=k_{l}^{2}\Theta _{l}(\theta )$, $l=0,1,\ldots $, which is
incorrect because $K$ is a two-variable operator. They compared the
resulting radial equation with that for a harmonic oscillator and concluded
that $k_{l}^{2}=l(l+1)$ and $E=\omega \left( 2n+l+\frac{3}{2}\right) $. BK
did not specify the values of $n$ but we may safely assume that $%
n=0,1,\ldots $. At this point it is worth noticing that BK's energies are
independent of $g_{1}$ which is most suspicious. In fact, the
Hellmann-Feynman theorem\cite{G32,F39} states that
\begin{equation}
\frac{dE}{dg_{1}^{2}}=\left\langle \frac{1}{\left( x_{1}+x_{2}-2x_{3}\right)
^{2}}\right\rangle >0.  \label{eq:HFT}
\end{equation}

In their section 4 BK appeared to solve the eigenvalue equation in a more
reasonable way starting from $F\Phi _{m}(\phi )=f_{m}^{2}\Phi _{m}(\phi )$, $%
m=0,1,\ldots $. Upon comparing their equations (4.4) and (4.5) they
concluded that $f_{m}=\lambda $ and $k_{l}^{2}=n(n+1)$. BK did not specify
the values of $n$ but it is clear that $n=-l-1$ or $n=l$ in order to be
consistent with the previous result for $k_{l}^{2}$. After some judicious
manipulation of their equations BK concluded that $g_{1}^{2}=3\left( \lambda
^{2}-\frac{1}{4}\right) $ from which we derive $f_{m}^{2}=\frac{1}{3}%
g_{1}^{2}+\frac{1}{4}$. In other words: given a value of the model parameter
$g_{1}$ all the eigenvalues of $F$ are equal, which does not seem to be
reasonable. Somewhat later BK obtained the expression $f_{m}=n+\frac{1}{2}$
which is intriguing because it relates the quantum numbers $m$ and $n$. From
all those equations we conclude that $g_{1}^{2}=3n(n+1)=3l(l+1)$ which makes
no sense because the model parameter $g_{1}$ (the strength of the Wolfes'
interaction) cannot depend on the quantum numbers. In the next section we
outline the solution of the eigenvalue equation for $H$ in spherical
coordinates in a more reasonable way.

\section{Correct solution in spherical coordinates}

\label{sec:correct_sol}

In order to solve the eigenvalue equation for $H$ in spherical coordinates
we first solve
\begin{equation}
F\Phi _{m}(\phi )=\left( -\frac{\partial ^{2}}{\partial \phi ^{2}}+\frac{%
g_{1}^{2}}{3\sin ^{2}\phi }\right) \Phi _{m}(\phi )=f_{m}^{2}\Phi _{m}(\phi
),\;m=0,1,\ldots ,  \label{eq:eigen_eq_Phi}
\end{equation}
and remove $\Phi _{m}(\phi )$ from the eigenvalue equation for $H$. Second,
we solve
\begin{equation}
\left( -\frac{1}{\sin \theta }\frac{\partial }{\partial \theta }\sin \theta
\frac{\partial }{\partial \theta }+\frac{f_{m}^{2}}{\sin ^{2}\theta }\right)
\Theta _{lm}(\theta )=k_{lm}^{2}\Theta _{lm}(\theta ),
\label{eq:eigen_eq_Theta}
\end{equation}
and remove $\Theta _{lm}(\theta )$ from the eigenvalue equation for $H$.
Finally, we are left with the radial equation
\begin{equation}
\left( -\frac{1}{2r^{2}}\frac{\partial }{\partial r}r^{2}\frac{\partial }{%
\partial r}+\frac{\omega ^{2}}{2}r^{2}+\frac{k_{lm}^{2}}{2r^{2}}\right)
R_{nlm}(r)=E_{nlm}R_{nlm}(r).  \label{eq:eigen_eq_r}
\end{equation}

It is not difficult to verify that the eigenvalues of this radial equation
are given by
\begin{equation}
E_{nlm}=\omega \left( 2n+s+\frac{3}{2}\right) ,\;s=\frac{1}{2}\left( \sqrt{%
k_{lm}^{2}+1}-1\right) ,  \label{eq:E_nlm}
\end{equation}
that resembles BK's equation (3.6), except for the fact that $s$ depends on $%
g_{1}$ as well as on the quantum numbers $l$ and $m$. In order to obtain the
exact analytical expression we have to solve equations (\ref{eq:eigen_eq_Phi}%
) and (\ref{eq:eigen_eq_Theta}). However it is not necessary, in our
opinion, because there is a much simpler approach outlined in the following
section.

\section{Exact textbook solution}

\label{sec:exact_sol}

The Hamiltonian operator in the center-of-mass and Jacobi coordinates (\ref
{eq:Jacobi_coord}) becomes

\begin{equation}
H=H_{d}-\frac{1}{2}\frac{\partial ^{2}}{\partial X^{2}},\;H_{d}=-\frac{1}{2}%
\sum_{i=1}^{3}\frac{\partial ^{2}}{\partial X_{i}^{2}}+\frac{\omega ^{2}}{2}%
\sum_{i=1}^{3}X_{i}^{2}+\frac{g_{1}^{2}}{6X_{2}^{2}}.  \label{eq:H_Jacobi}
\end{equation}
We appreciate that the spectrum of $H$ is continuous while that of $H_{d}$
is discrete. In fact, in the two preceding sections we discussed the
solutions to the eigenvalue equation for $H_{d}$ in spherical coordinates.
It is clear that $H_{d}$ is separable in Jacobi coordinates
\begin{eqnarray}
H_{d} &=&H_{1}+H_{2}+H_{3},  \nonumber \\
H_{i} &=&-\frac{1}{2}\frac{\partial ^{2}}{\partial X_{i}^{2}}+\frac{\omega
^{2}}{2}X_{i}^{2},\;i=1,3,  \nonumber \\
H_{2} &=&-\frac{1}{2}\frac{\partial ^{2}}{\partial X_{2}^{2}}+\frac{\omega
^{2}}{2}X_{2}^{2}+\frac{g_{1}^{2}}{6X_{2}^{2}}.  \label{eq:H_d_Jacobi}
\end{eqnarray}
\newline
Note that $H_{1}$ and $H_{3}$ are one-dimensional harmonic oscillators (HO),
whereas $H_{2}$ is the so-called singular harmonic oscillator (SHO).
Therefore, if we try a solution of the form
\begin{equation}
\psi _{n_{1}n_{2}n_{3}}\left( X_{1},X_{2},X_{3}\right) =\varphi
_{n_{1}}^{HO}\left( X_{1}\right) \varphi _{n_{2}}^{SHO}\left( X_{2}\right)
\varphi _{n_{3}}^{HO}\left( X_{3}\right) ,  \label{eq:Psi_Jacobi}
\end{equation}
we conclude that
\begin{eqnarray}
H_{i}\varphi _{n_{i}}^{HO} &=&E_{n_{i}}^{HO}\varphi
_{n_{i}}^{HO},\;i=1,3,\;H_{2}\varphi _{n_{2}}^{SHO}=E_{n_{2}}^{SHO}\varphi
_{n_{2}}^{SHO},  \nonumber \\
E_{n_{i}}^{HO} &=&\omega \left( n_{i}+\frac{1}{2}\right) ,\;n_{i}=0,1,\ldots
,  \nonumber \\
E_{n_{1}n_{2}n_{3}} &=&E_{n_{1}}^{HO}+E_{n_{2}}^{SHO}+E_{n_{3}}^{HO}.
\label{eq:E_n1n2n3}
\end{eqnarray}
The eigenvalues of the singular harmonic oscillator can also be calculated
analytically\cite{PR03}:

\begin{equation}
E_{n_{2}}^{SHO}=\omega \left( 2n_{2}+\frac{1}{2}+\sqrt{\frac{1}{4}+\frac{%
g_{1}^{2}}{3}}\right) .  \label{eq:E^SHO}
\end{equation}

\section{Conclusions}

\label{sec:conclusions}

The main conclusion of this paper is that the problem can be solved more
easily and straightforwardly in Jacobi coordinates which makes it
unnecessary to resort to spherical ones. However, we have outlined an
earlier solution of the problem in spherical coordinates in order to compare
those and present results.  In section~\ref{sec:correct_sol} we outlined a
more rigorous and clearer solution in terms of spherical coordinates and in
section~\ref{sec:exact_sol} the simpler procedure in terms of Jacobi
coordinates proposed here.

\end{document}